\documentclass[twocolumn,showpacs,preprintnumbers,amsmath,amssymb,nofootinbib]{revtex4}
\input psfig.sty
\usepackage{epsf}
\usepackage{graphicx}  
\usepackage{dcolumn}   
\usepackage{bm}        

\newcommand{\be}{\begin{equation}}
\newcommand{\en}{\end{equation}}
\newcommand{\bea}{\begin{eqnarray}}
\newcommand{\ena}{\end{eqnarray}}

\begin{document}

\title{Latest supernovae constraints on $f(R)$ cosmologies} 

\author{J. Santos$^1${\footnote{janilo@dfte.ufrn.br}}}


\author{J. S. Alcaniz${^{2,3}}${\footnote{alcaniz@on.br}}}

\author{F. C. Carvalho$^4${\footnote{fabiocc@das.inpe.br}}}

\author{N. Pires$^1${\footnote{npires@dfte.ufrn.br}}}

\affiliation{$^1$Universidade Federal do Rio Grande do Norte, Departamento de
F\'{\i}sica, 59072-970 Natal - RN, Brasil}


\affiliation{$^2$Observat\'orio Nacional, 20921-400 Rio de Janeiro - RJ, Brasil}

\address{$^3$Instituto Nacional de Pesquisas Espaciais/CRN, 59076-740, Natal - RN, Brasil}

\affiliation{$^4$Instituto Nacional de Pesquisas Espaciais, 12227-010, S\~ao Jos\'e dos Campos - SP, Brasil}

\date{\today}


\begin{abstract}
A class of modified gravity, known as $f(R)$-gravity, has presently been applied to Cosmology as a realistic alternative to dark
energy. In this paper we use the most recent Type-Ia Supernova (SNe Ia) data, the so-called \emph{Union} sample of 307 SNe Ia, to place bounds on a theory of the form $f(R)=R - \beta/R^n$ within the Palatini approach.  Given the complementarity of SNe Ia data with other cosmological observables, a joint analysis with  measurements of baryon acoustic oscillation peak and estimates of the {\rm CMB} shift parameter is also performed. We show that, for the allowed intervals of $n$, $\Omega_{mo}$, and $\beta$, models based on $f(R) = R - \beta/R^{n}$ gravity in the Palatini approach can produce the sequence of radiation-dominated, matter-dominated, and accelerating periods without need of dark energy.

\end{abstract}


\maketitle


\section{Introduction}


One of the key problems at the interface between fundamental physics and cosmology is to understand the physical mechanism behind the late-time acceleration of the Universe. In principle, this phenomenon may be the result of unknown physical processes involving either modifications of gravitation theory or the existence of new fields in high energy physics. Although the latter route is most commonly used, which gives rise to the idea of a dark energy component (see, e.g., \cite{alc}), following the former, at least two other attractive approaches to this problem can be explored. The first one is related to the possible existence of extra dimensions, an idea that links cosmic acceleration with the hierarchy problem in high energy physics, and gives rise to the so-called brane-world cosmology \cite{RandSund1999}. The second one, known as $f(R)$ gravity, examine the possibility of modifying Einstein's general relativity (GR) by adding terms proportional to powers of the Ricci scalar $R$ to the Einstein-Hilbert Lagrangian \cite{Buchdahl}. The cosmological interest in $f(R)$ gravity comes from the fact that these theories can exhibit naturally an accelerating expansion without introducing dark energy.  However, the freedom in the choice of different functional forms of $f(R)$ gives rise to the problem of how to constrain on theoretical and/or observational grounds, the many possible $f(R)$ gravity theories.
Much efforts within the realm, mainly from a theoretical viewpoint, have been developed so far~\cite{Amendola} (see also Refs.~\cite{Francaviglia} for recent reviews), while only recently observational constraints from several cosmological data sets have been explored for testing the viability of these theories\cite{Amarzguioui,Tavakol,Fairbairn,Movahed,Fabiocc,tv}.

An important aspect worth emphasizing concerns the two different variational approaches that may be followed when one works with $f(R)$ gravity theories, namely, the metric  and the Palatini formalisms (see, e.g., \cite{Francaviglia}). In the metric formalism the connections are assumed to be the Christoffel symbols and variation of the action is taken with respect to the metric, whereas in the Palatini variational approach the metric and the affine connections are treated as independent fields and the variation is taken with respect to both. In fact, these approaches are equivalents only in the context of GR, that is, in the case of linear Hilbert action; for a general $f(R)$ term in the action they give different equations of motion.

In the present paper we will restrict ourselves to the Palatini formalism for gravitation and will focus on its application to the flat Friedmann-Robertson-Walker (FRW) cosmological model.  We will derive constraints on the two parameters $n$ and $\beta$ of the
$f(R) = R - \beta/R^{n}$  gravity theory from the most recent compilations of type Ia Supernovae (SNe Ia) observations, which includes the
recent large samples from
Supernova Legacy Survey (SNLS), the ESSENCE Survey, distant SNe Ia observed with HST, and others, giving a sample of 307 SNe Ia events \cite{0804.4142}. We also combine the SNe Ia data with information from the baryon acoustic oscillation ({\rm BAO}) \cite{Eisenstein2005} and the {\rm CMB} shift parameter \cite{wmap} in order to improve the SNe Ia bounds on the free parameters of the theory.

\section{Palatini Approach}

The action that defines an $f(R)$ gravity is given by
\begin{equation}
\label{actionJF}
S = \frac{1}{2\kappa^2}\int d^4x\sqrt{-g}f(R) + S_m\,,
\end{equation}
where $\kappa^2=8\pi G$, $g$ is the determinant of the metric tensor and $S_m$ is the standard action for the matter
fields. Treating the metric and the connection as completely independent fields, variation of this action gives the field equations
\begin{equation}
\label{field_eq}
f'R_{(\mu\nu)} - \frac{f}{2}g_{\mu\nu}  = \kappa^2T_{\mu\nu}\,,
\end{equation}
where $T_{\mu\nu}$ is the matter energy-momentum tensor which, for a perfect-fluid, is given by
$T_{\mu\nu} = (\rho_m + p_m)u_{\mu}u_{\nu} + p_m g_{\mu\nu}$,
where $\rho_m$ is the energy density, $p_m$ is the fluid pressure and $u_{\mu}$
is the fluid four-velocity. Here, we adopt the notation $f'=df/dR$,
$f''=d^2f/dR^2$ and so on. In (\ref{field_eq}) $R_{\mu\nu}$ is given in the usual way in terms of the independent connection $\Gamma_{\mu\nu}^{\rho}$, and its derivatives, which is related with the Christoffel symbol $\left\{^{\rho}_{\mu\nu}\right\}$ of the metric $g_{\mu\nu}$ by
\begin{equation}
\Gamma_{\mu\nu}^{\rho} = \left\{^{\rho}_{\mu\nu}\right\} + \frac{1}{2f'}\left( \delta^{\rho}_{\mu}\partial_{\nu} +
\delta^{\rho}_{\nu}\partial_{\mu} - g_{\mu\nu}g^{\rho\sigma}\partial_{\sigma} \right)f'
\end{equation}
and $R=g^{\mu\nu}R_{\mu\nu}$.

We assume a homogeneous and isotropic FRW universe whose metric is
$g_{\mu\nu}=diag(-1,a^2,a^2,a^2)$, where $a(t)$ is the cosmological scale factor.
The generalized Friedmann equation can be written in terms of redshift parameter $z=a_0/a -1$ and the density parameter $\Omega_{mo} \equiv \kappa\rho_{mo}/(3H_0^2)$ as (see Ref.~\cite{Tavakol,Fabiocc} for details)
\begin{equation}
\label{fe3}
\frac{H^2}{H_0^2} = \frac{3\Omega_{mo}(1 + z)^3 + f/H_0^2}{6f'\xi^2}\,,
\end{equation}
where
\begin{equation}
\label{xi}
 \xi = 1 + \frac{9}{2}\,\frac{f''}{f'}\,\frac{H_0^2\Omega_{mo}(1+z)^3}{Rf'' - f'}
\end{equation}
and $\rho_{mo}$ is the matter density today. The trace of Eq. (\ref{field_eq}) gives another relation
\begin{equation}
\label{trace2}
Rf' - 2f = -3H_0^2\Omega_{mo}(1 + z)^3\,,
\end{equation}
and, as can be easily checked, for the Einstein-Hilbert Lagrangean ($f=R$) Eq. (\ref{fe3}) reduces to the known form of Friedmann equation.

By assuming a functional form of the type $f(R) = R - \beta/R^{n}$, 
one may easily show that Eq. (\ref{trace2})  evaluated at $z=0$ imposes the following relation among $n$, $\Omega_{mo}$ and $\beta$
\begin{equation}
\beta = \frac{R_0^{n+1}}{n+2}\,\left( 1 - \frac{3\Omega_{mo}H^2_0}{R_0} \right)\,,
\end{equation}
where $R_0$, the value of the Ricci scalar today, is determined from the algebraic equation resulting from equating (\ref{trace2}) and (\ref{fe3}) for $z=0$. Hence, specifying the values of two of these parameters the third is automatically fixed. In other words, in the Palatini approach, the two parameter of $f(R)=R-\beta/R^{n}$ can be thought as the pair ($n,\beta$) or ($n,\Omega_{mo}$).

\begin{figure}
\centerline{\psfig{figure=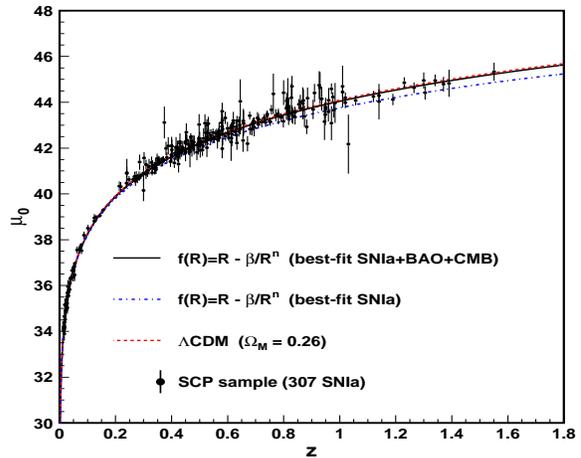,width=3.3truein,height=2.7truein}
\hskip 0.1in}
\caption{Hubble diagram for 307 SNe Ia from the \emph{Union} sample \cite{0804.4142}. The curves correspond to the best-fit pairs of $n$ and $\Omega_{mo}$ arising from statistical analyses involving SNe Ia (only) and SNe Ia + BAO + CMB shift parameter. For the sake of comparison the flat $\Lambda$CDM scenario with $\Omega_{mo} = 0.26$ is also shown.}
\label{figh}
\end{figure}

\begin{figure*}
\centerline{\psfig{figure=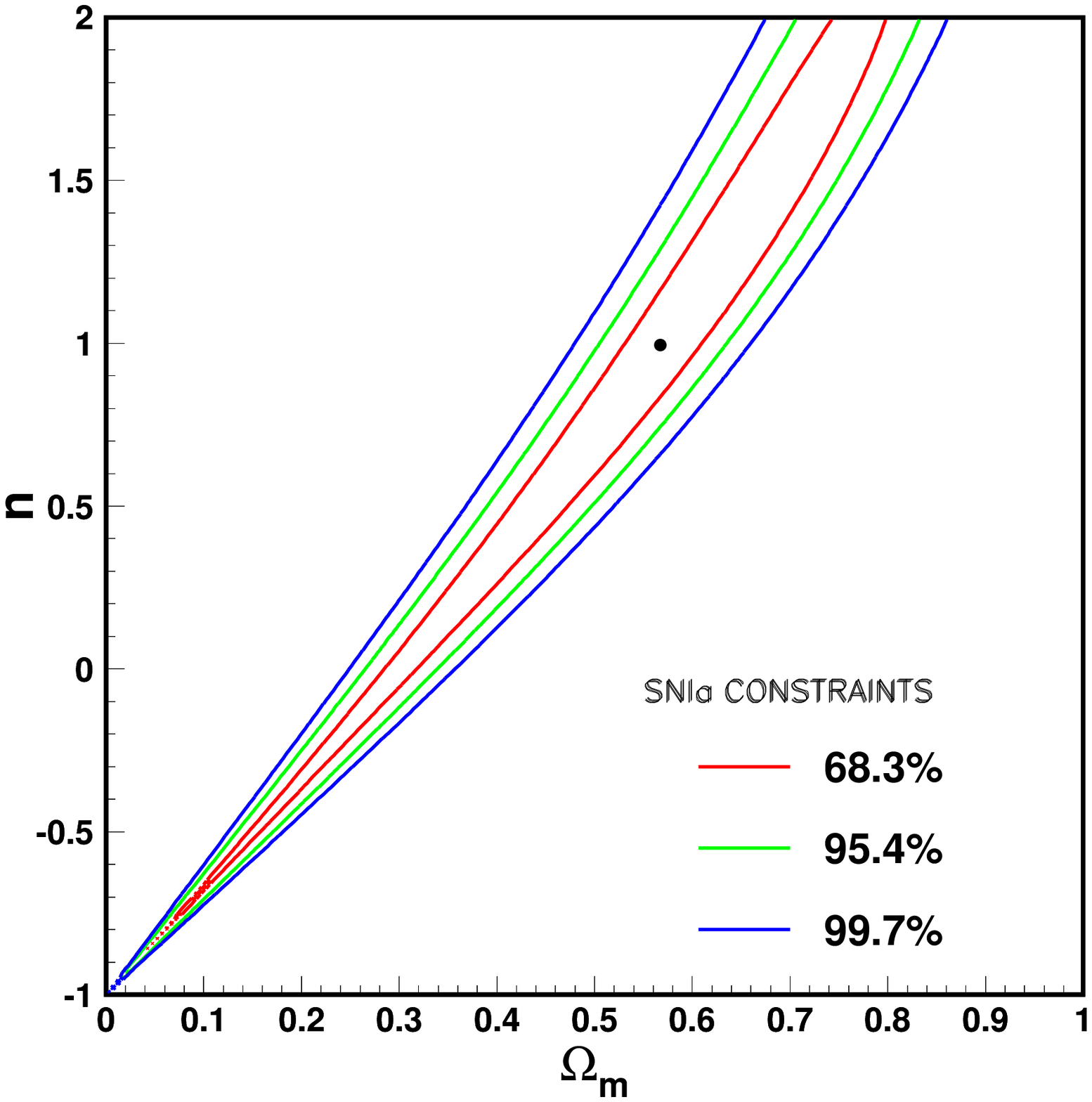,width=2.3truein,height=2.7truein}
\psfig{figure=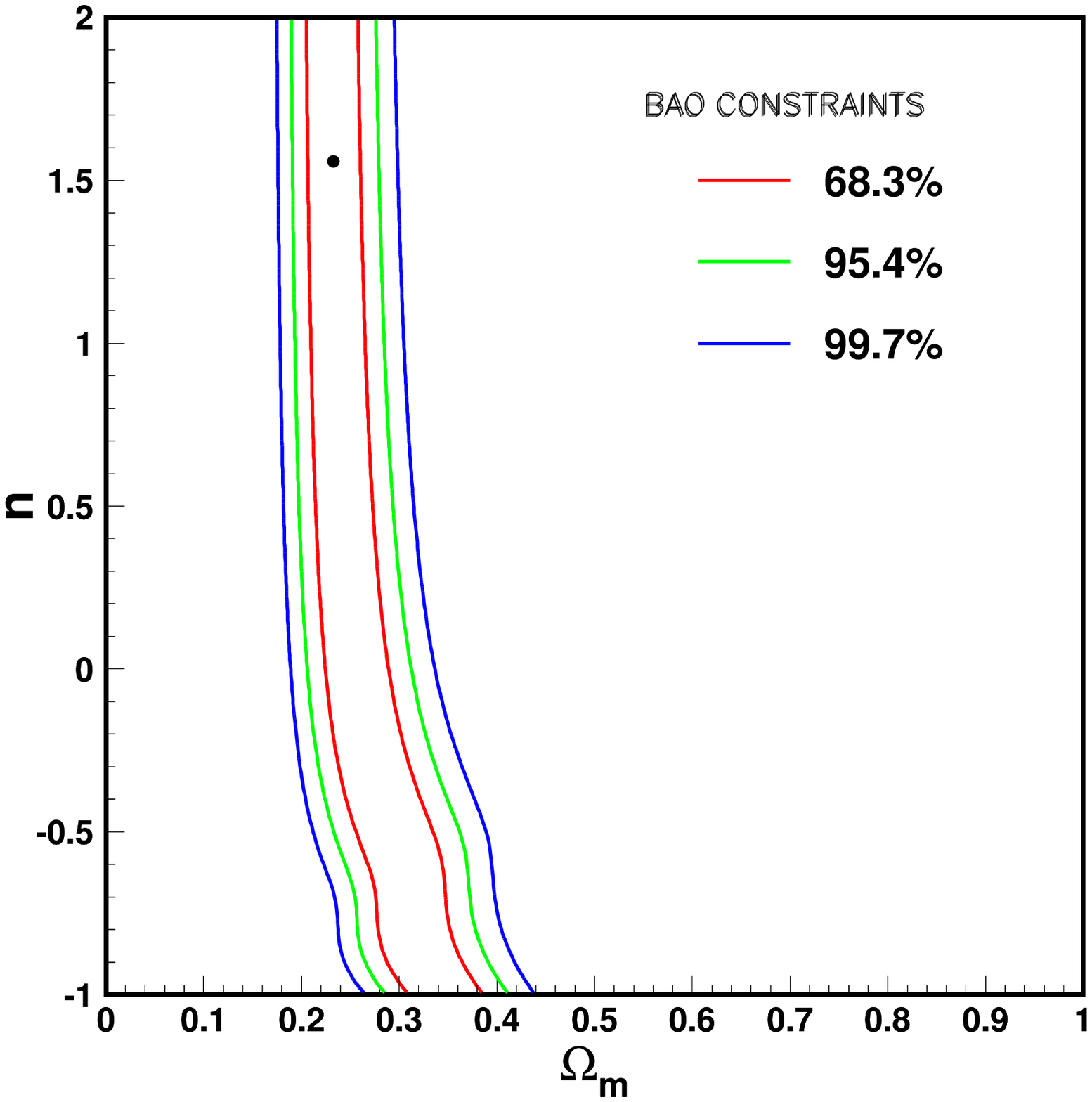,width=2.3truein,height=2.7truein}
\psfig{figure=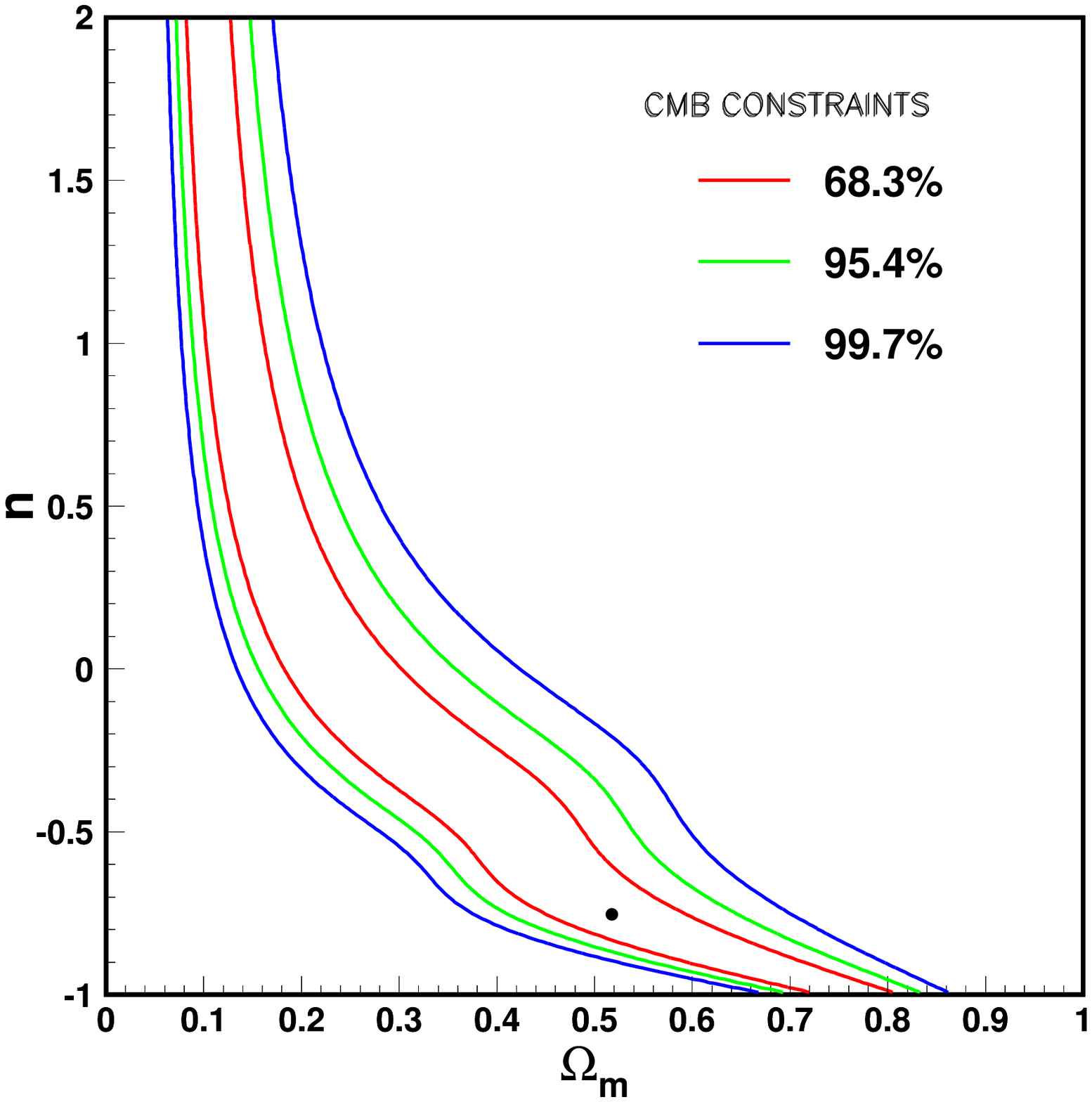,width=2.3truein,height=2.7truein}
\hskip 0.1in}
\caption{Confidence intervals at 68.3\%, 95.4\% and 99.73\% in the $n \times \Omega_{mo}$ plane arising from: {\bf{Left)}} SNe Ia (\emph{Union} sample); {\bf{Middle)}} BAO (SDSS); {\bf{Right)}} CMB shift parameter (WMAP5). In all panels, the dot marks the best-fit pair ($n$, $\Omega_{mo}$) for each analysis. Note that the contours of the allowed parameters from these measurements are roughly complementary in the $n \times \Omega_{mo}$ space.}
\label{figh}
\end{figure*}

\section{Observational analyses}

Since the very first results showing direct evidence for a present cosmic acceleration (using a small number of SNe Ia events) \cite{perl}, the number and quality of SNe Ia data available for cosmological studies have increased considerably due to several observational programs. The most up to date set of SNe Ia has been compiled by Kowalski {\it et al.} \cite{0804.4142} and includes recent large samples from SNLS \cite{Astier} and ESSENCE \cite{essence} surveys, older data sets and the recently extended data set of distant supernovae observed with HST. The total compilation, the so-called \emph{Union} sample, amounts to 414 SNe Ia events, which was reduced to 307 data points after selection cuts.

In this Section, we will use this SNe Ia sample to place limits on the $n - \Omega_{mo}$ (or, equivalently, $\beta - \Omega_{mo}$) parametric space. This analysis, therefore, updates the results of Refs.~\cite{Amarzguioui,Tavakol}. We also perform a joint analysis involving the \emph{Union} SNe Ia sample and measurements of the  baryonic acoustic oscillations (BAO) from SDSS \cite{Eisenstein2005} and the CMB shift parameter as given by the WMAP team \cite{wmap} to break possible degeneracies in the $n - \Omega_{mo}$ plane (for more  details on the statistical analyses discussed below we refer the reader to Ref.~\cite{refs}).


\subsection{Latest SNe Ia constraints}


The predicted distance modulus for a supernova at redshift $z$, given a set of
parameters $\mathbf{P} = (n, \Omega_{mo})$, is
\begin{equation} \label{dm}
\mu_0(z|\mathbf{P}) = m - M = 5\,\mbox{log} d_L + 25,
\end{equation}
where $m$ and $M$ are, respectively, the apparent and absolute magnitudes, and $d_L$ stands for the luminosity distance (in units of megaparsecs),
\begin{equation}
\label{LuminosityDistance}
d_L(z;{\bf P})=(1+z)\int_{0}^{z}\frac{dz^{\prime}}{H(z^{\prime};{\bf P})}\;,
\end{equation}
where $H(z; {\bf P})$ is given by Eqs. (\ref{fe3}) - (\ref{trace2}).

We estimate the best fit to the set of parameters $\mathbf{P}$ by using a $\chi^{2}$ statistics, with
\begin{equation} \label{chi2307}
\chi^{2}_{SNe} = \sum_{i=1}^{N}{\frac{\left[\mu_0^{i}(z|\mathbf{P}) -
\mu_{obs}^{i}(z)\right]^{2}}{\sigma_i^{2}}},
\end{equation}
where $\mu_p^{i}(z|\mathbf{P})$ is given by Eq. (\ref{dm}), $\mu_o^{i}(z)$ is the extinction corrected distance modulus for a given SNe Ia at $z_i$, and $\sigma_i$ is the uncertainty in the individual distance moduli. Since we use in our analysis the \emph{Union} sample (see \cite{0804.4142} for details),  $N = 307$.

Figure (1) shows the Hubble diagram for the 307 SNe Ia events of the \emph{Union} sample. The curves stand for the best-fit $f(R)$ models obtained from SNe Ia and SNe Ia + BAO + CMB analysis. For the sake of comparison, the standard $\Lambda$CDM model with $\Omega_{mo} = 0.26$ is also shown. Note that all models seem to be able to reproduce fairly well the SNe Ia measurements. In  Fig. (2a) we show the first results of our statistical analyses. Contour plots (68.3\%, 95.4\% and 99.7\% c.l.) in the $n\times\Omega_{mo}$ plane are shown for the $\chi^2$ given by Eq. (\ref{chi2307}). We clearly see that SNe Ia measurements alone do not tightly constrain  the values of $n$ and $\Omega_{mo}$, allowing for a large interval of values for these parameters, with $n$ ranging from -1 to even beyond 1, and $\Omega_{mo}$ consistent with both vacuum solutions ($\Omega_{mo} = 0$), as well with universes with up to 90\% of its energy density in the form of non-relativistic matter. The best-fit values for this analysis are $\Omega_{mo}=0.57$ and $n=0.99$, with the reduced $\chi^2_r \equiv \chi^2_{min}/\nu \simeq 1.01$ ($\nu$ is defined as degrees of freedom).

\begin{figure*}
\centerline{\psfig{figure=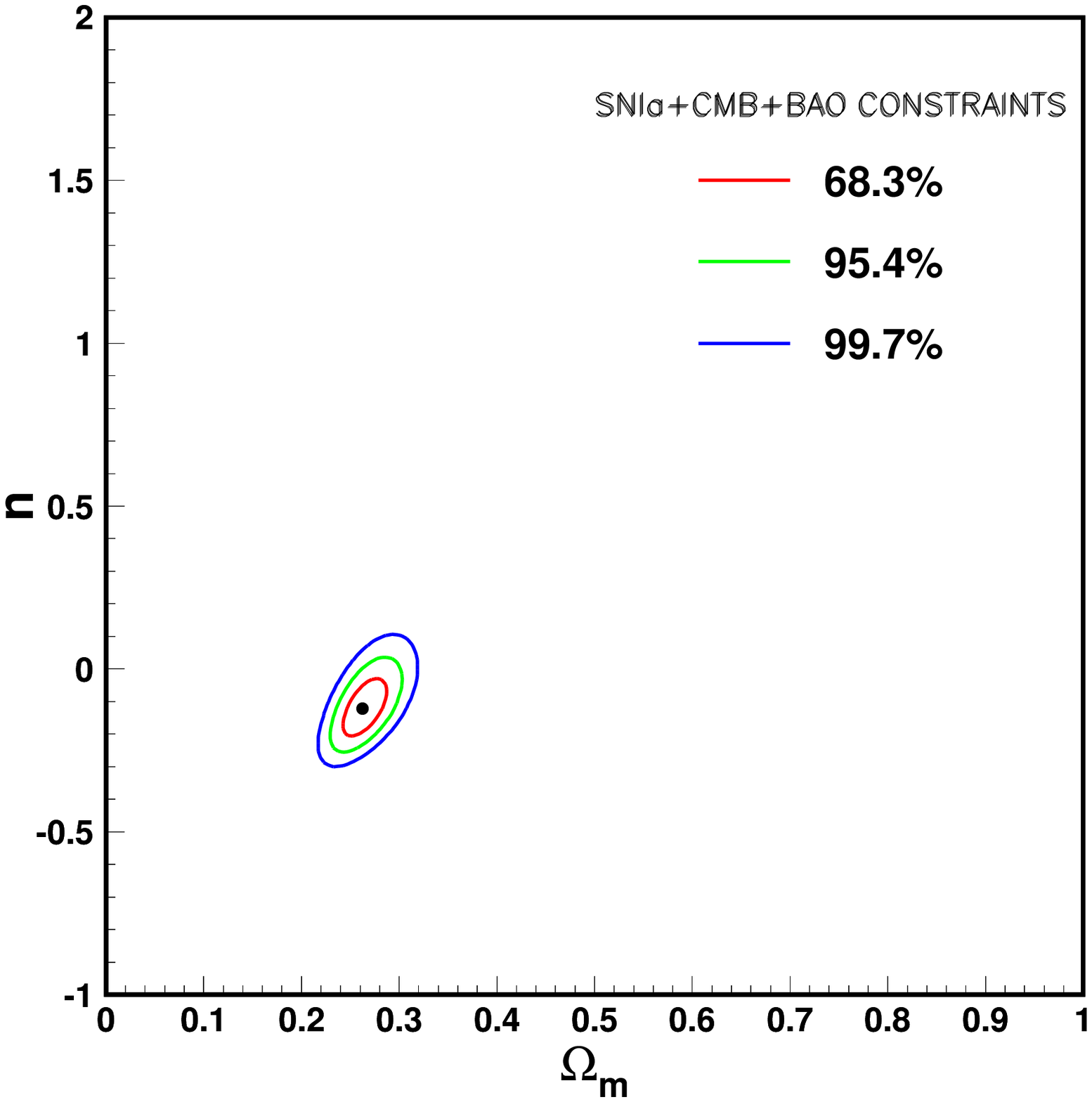,width=3.3truein,height=2.8truein}
\hskip 0.4in
\psfig{figure=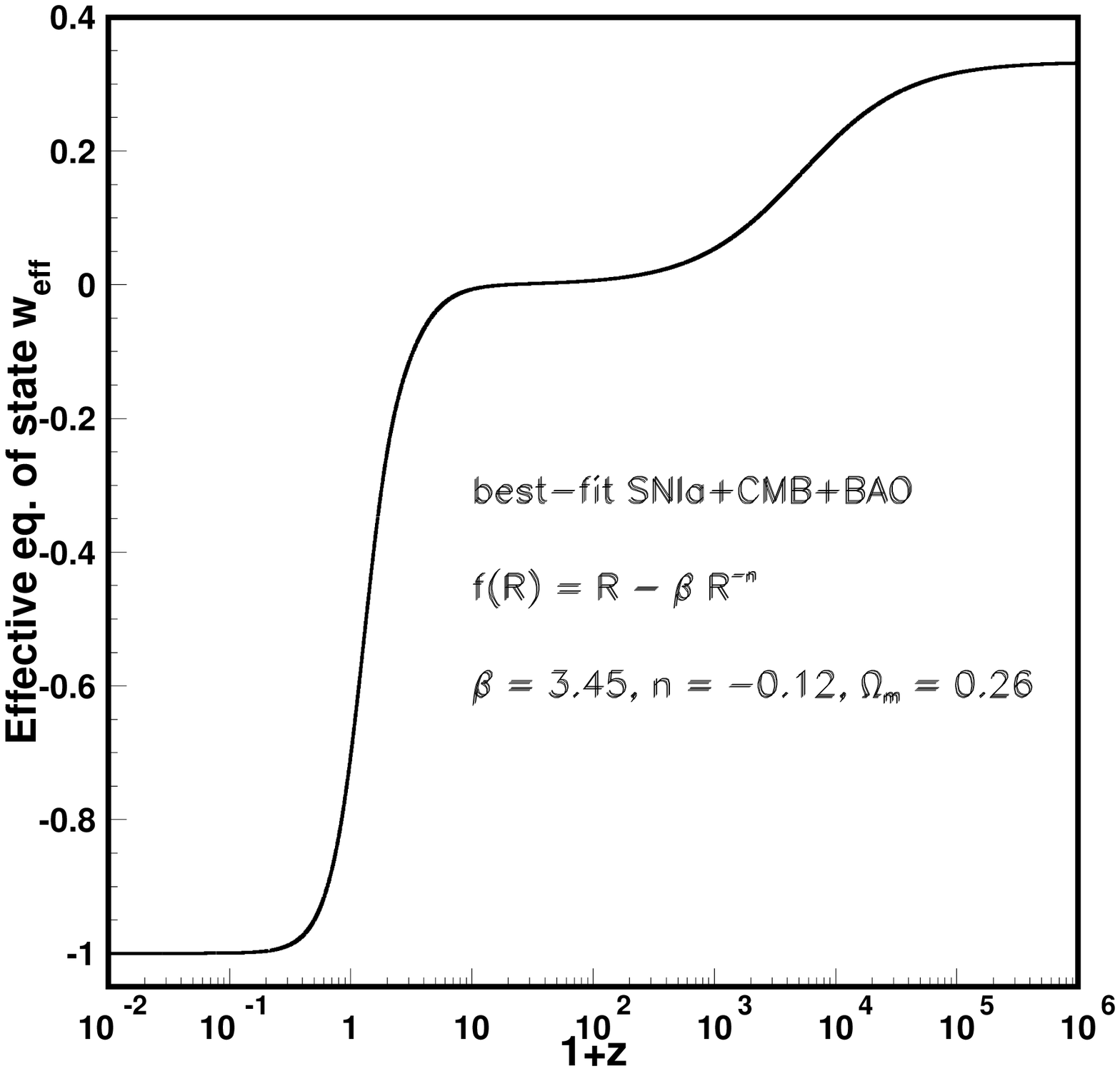,width=3.3truein,height=2.8truein}
\hskip 0.1in}
\caption{{\bf{Left:}} Confidence intervals at 68.3\%, 95.4\% and 99.73\% in the $n \times \Omega_{mo}$ plane arising from the combined fit involving SNe Ia \emph{Union} sample, BAO and CMB shift parameter. The best-fit values for this analysis is $n = -0.12$ and $\Omega_{mo} = 0.26$ ($\beta = 3.45$). {\bf{Right:}} Effective equation of state as a function of redshift for a $f(R)=R-\beta/R^{n}$ theory in the Palatini formalism. The parameters values correspond to the best-fit solution of our joint analysis.}
\label{figh}
\end{figure*}

\subsection{Joint analysis}

\subsubsection{BAO}

The acoustic oscillations of baryons in the primordial plasma leave a signature on the correlation function of galaxies as observed by Eisenstein {\it et al.} \cite{Eisenstein2005}. This signature furnishes a standard rule which can be used to constrain the following quantity:
\begin{equation}
{\cal{A}} = D_V\frac{\sqrt{\Omega_{mo} H_0^2}}{z_*}\;,
\end{equation}
where the observed value is ${\cal{A}}_{obs} = 0.469 \pm 0.017$, $z_* = 0.35$ is the typical redshift of the SDSS sample and $D_V$ is the dilation scale, defined as
$D_V = \left[{D_M}(z_*|{\bf P})^2{z_*}/{H(z_*|{\bf P})}\right]^{1/3}$
with the comoving distance $D_M$ given by
$D_M(z_*|{\bf P}) = \int_{0}^{z_*}{{dz'}/{H(z'|{\bf P})}}$. In Fig. (2b) we show the confidence contours (68.3\%, 95.4\% and 99.7\% c.l.) in the $n - \Omega_{mo}$ plane arising from this measurement of ${\cal{A}}$. As expected, since this quantity has been measured at a specific redshift ($z_{*} = 0.35$), it forms bands on this parametric space, instead of ellipsoids as in the case of SNe Ia data.

\subsubsection{Shift Parameter}

The shift parameter  ${\mathcal{R}}$, which determines the whole shift of the CMB angular power spectrum, is given by \cite{b}
\begin{equation}
\label{ShiftParameter}
\mathcal{R}\equiv\sqrt{\Omega_{mo}}\int_0^{z_{\mathrm ls}}\frac{H_0\,dz'}{H(z';{\bf P})},
\end{equation}
where the $z_{\mathrm ls} = 1089$ is the redshift of the last  scattering surface, and the current estimated value for this quantity is ${\mathcal R}_{\mathrm{obs}} = 1.70\pm 0.03 $ \cite{wmap}. Note that, to include the CMB shift parameter into the analysis, the equations of motion must be integrated up to the matter/radiation decoupling, $z \simeq 1089$. Since radiation is no longer negligible at this redshift, a radiation component with
an energy density today of $\Omega_{\gamma}=5\times 10^{-5}$ has been included in our analysis. Figure (2c) shows the constraints on the $n - \Omega_{mo}$ plane from the current WMAP estimate of ${\mathcal{R}}$.

\subsubsection{Results}

In Fig. (3a) we show the results of our joint SNe Ia + BAO + CMB analysis. Given the complementarity of these measurements in the $n - \Omega_{mo}$ plane [see Figure (2)], we obtain a considerable enhancement of the constraining power over $n$ and $\Omega_{mo}$ from this combined fit. Note also that the best-fit value for the matter density parameter, i.e., $\Omega_{mo}=0.26$, is consistent with current estimates of the contribution of non-relativistic matter to the total energy density in the universe (see, e.g., \cite{wmap}). The joint fit also constrains the parameters $n$, $\Omega_{mo}$, and $\beta$ to lie in the following intervals (at 99.7\% c.l.)
\[
n\in [-0.3, 0.1], \quad \Omega_{mo}\in [0.22, 0.32] \quad \mbox{and} \quad \beta\in [1.3, 5.5],
\]
which is consistent with the results obtained in Refs. \cite{Amarzguioui,Tavakol} using the supernova \emph{Gold} and the SNLS data sets, respectively.
\begin{table}[h]
\begin{center}
\begin{tabular}{lcrl}
\hline \hline \\
Test& Ref. & $n$ & $\beta$\\
\hline \hline \\
SNe Ia ({\emph{Gold}}) & \cite{Amarzguioui} & 0.51 & 10\\
SNe Ia ({\emph{Gold}}) + BAO + CMB & \cite{Amarzguioui} & -0.09 & 3.60\\
SNe Ia (SNLS) & \cite{Tavakol} & 0.6 & 12.5\\
SNe Ia (SNLS) + BAO + CMB & \cite{Tavakol} & 0.027 & 4.63\\
H(z) & \cite{Fabiocc} & -0.90 & 1.11\\
H(z) + BAO + CMB& \cite{Fabiocc} & 0.03 & 4.70\\
LSS & \cite{tv} & 2.6 & -\\
SNe Ia (\emph{Union}) & This Paper & 0.99 & -\\
CMB & This Paper & -0.75 & 0.48\\
BAO & This Paper & 1.56 & -\\
SNe Ia (\emph{Union}) + BAO + CMB& This Paper & -0.12 & 3.45\\
\hline \hline \\
\end{tabular}
\end{center}
\caption{Best-fit values for $n$ and $\beta$ (the $\Lambda$CDM model corresponds to $n = 0$ and $\beta = 4.38$).}
\end{table}

\subsection{Effective equation of state}

Recently, Amendola {\it et al.} \cite{Amendola-b} showed that $f(R)$ derived cosmologies in the metric formalism cannot produce a standard matter-dominated era followed by an accelerating expansion. To verify if the same undesirable behavior also happens in the Palatini formalism adopted in this paper, we first derive the effective equation of state (EoS)
\begin{equation}
w_{eff} = -1 + \frac{2(1+z)}{3H}\frac{dH}{dz}
\end{equation}
as a function of the redshift.

Figure (3b) shows the effective EoS as a function of $1 + z$ for the best-fit solution of our joint SNe Ia + BAO + CMB analysis. Note that, for this particular combination of parameters, the universe goes through the last three phases of cosmological evolution, i.e., radiation-dominated ($w=1/3$), matter-dominated ($w=0$) and the late time acceleration phase (in this case with $w\simeq -1$). Therefore, the arguments of Ref.~\cite{Amendola-b} about the $w_{eff}$ in the metric approach seem not to apply to the Palatini formalism, at least for the interval of parameters $n$, $\Omega_{mo}$ and $\beta$ given by our statistical analysis. In Table I we summarize the main results of this paper compare them with recent determinations of the parameters $n$ and $\beta$ from independent analyses.


\section{Conclusions}


$f(R)$-gravity based cosmology has presently been thought of as a realistic alternative to general relativistic dark energy models. In this paper, we have worked in the context of a $f(R) = R - \beta R^{-n}$ gravity with equations of motion derived according to the Palatini approach. We have performed consistency checks and tested the observational viability of these scenarios by using the latest sample of SNe Ia data, the so-called \emph{Union} sample of 307 events. Although the current SNe Ia measurements alone cannot constrain significantly the model
parameters $n$, $\Omega_{mo}$ and $\beta$, when combined with information from BAO and CMB shift parameters, the fit leads to very restrictive constraints on the $n - \Omega_{mo}$ (or, equivalently, $\beta - \Omega_{mo}$) parametric space. At 99.7\% c.l., e.g., we have found the intervals $n\in [-0.25, 0.35]$ and $\Omega_{mo} \in [0.2, 0.31]$ ($\beta\in [2.3, 7.1]$). We note that, differently from results in the metric formalism \cite{Amendola-b}, the universe corresponding to the best-fit solution for a combined SNe Ia+BAO+CMB $\chi^2$ minimization ($\Omega_{mo} = 0.26$ and $n = -0.12$) shows all three last phases of the cosmological evolution: radiation era, matter era and a late time cosmic acceleration.

\acknowledgements

The authors are very grateful to Edivaldo Moura Santos for helpful conversations and a critical reading of the manuscript. JS and NP thank financial support from PRONEX (CNPq/FAPERN). FCC acknowledge financial support from FAPESP. JSA's work is supported by CNPq.

\end{document}